\documentclass[twocolumn,showpacs,preprintnumbers,amsmath,amssymb]{revtex4}
\pdfoutput=1

\usepackage{graphicx}
\usepackage{dcolumn}
\usepackage{bm}

\begin{document}


\title{Harmonic Measure for Percolation and Ising Clusters Including Rare Events}

\author{David A. Adams, (1) Leonard M. Sander, (1 and 2) and Robert M. Ziff (2 and 3)}
\date{\today}

\begin{abstract}
We obtain the harmonic measure of the hulls of critical percolation clusters and Ising-model Fortuin-Kastelyn clusters using a biased random-walk sampling technique which allows us to measure probabilities as small as $10^{-300}$.  We find the multifractal $D(q)$ spectrum including regions of small and negative $q$.  Our results for external hulls agree with Duplantier's theoretical predictions for $D(q)$ and his exponent $-23/24$ for the harmonic measure probability distribution.  For the complete hull, we find the probability decays with an exponent of $-1$ for both systems.
\end{abstract}

\address{$^{1}$Department of Physics, University of Michigan, Ann Arbor MI 48109-1040 \\
$^{2}$Michigan Center for Theoretical Physics, University of Michigan, Ann Arbor MI 48109-1040
\\$^{3}$Department of Chemical Engineering, University of Michigan, Ann Arbor MI 49109-2136}

\maketitle


The  harmonic measure is a fundamental property of  geometric objects, and its determination is of considerable interest.   For a given object, it may be defined in the following way: consider the object to be a grounded conductor with fixed charge of unity. Then the harmonic measure, $\mu$, is the distribution of electric field on the hull (surface) of the object. Alternatively, we may imagine that we allow many random walkers to start far away from the object and record where they land. The probability density of hitting the hull at a point is $\mu$. This quantity is of both theoretical \cite{Meakin,Duplantier_Per} and practical \cite{Halsey92b} interest. In particular, if the shape in question is  fractal, $\mu$  shows interesting scaling properties.

In this Letter we show how to find $\mu$ numerically for two fundamental  systems that produce fractal clusters  in two dimensions, namely percolation \cite{Stauffer94} and  Fortuin-Kasteleyn (FK) \cite{KF69} 
clusters in the Ising model.  Our method allows us to sample very small probabilities (of order $10^{-300}$) using random walker simulations. In the case of these fractals, $\mu$ is of particular theoretical importance because it is \emph{multifractal}.  Our large dynamic range allows us to explore this property fully.


The harmonic measure $\mu$ is non-negative and normalized on the hull: $\int d \mu = 1$.  A partition function \cite{Halsey86} can be defined by dividing the hull into $j$ boxes of length $l$,
\begin{equation}
Z_q = \sum_j p_j^q
\label{eq:Zq},
\end{equation}
where $p_j = \int d \mu$ over box $j$.  For large fractal clusters  $Z_q$ scales as a power-law,
\begin{equation}
Z_q \approx \left( { l\over R} \right)^{-\tau(q)} 
\label{eq:Zq2},
\end{equation}
where $R$ is the length scale of the cluster and $$\tau(q) \equiv (q-1) D(q)$$ is the multifractal scaling exponent; $D(q)$ is called the generalized dimension.  
We recall some special values of $D(q)$: $D(0)$ is the fractal dimension of the support of the measure, which describes the region the hull covers.  
Additionally, $D(1) = 1$ is known from Makarov's theorem \cite{Makarov}. A related quantity of interest is the  curve $f(\alpha)$, which is the Legendre transform of $\tau(q)$:
\begin{equation}
f(\alpha) = q{d \tau \over d q}- \tau, \quad  \alpha = {d \tau \over d q}
\label{eq:falpha}.
\end{equation}
The rest of this paper will focus on $\tau(q)$ and $D(q)$. We are also able to produce $f(\alpha)$ by numerically taking the Legendre transform of our measured $\tau(q)$.

The exact spectrum of $D(q)$ has been obtained for percolation \cite{Duplantier_Per} and the more general $Q$-state Potts model \cite{Duplantier_Potts}, derived from generalized conformal invariance in terms of a central charge $c$,
\begin{equation}
\label{eq:Dc}
D(q) = {1 \over 2} + \left( \sqrt{ {24q+1-c \over 25-c}} -1\right)^{-1} \quad 
q \in [-{1-c\over 24}, +\infty) 
.
\end{equation}
Percolation and FK clusters for the Ising model correspond  to $c=0$ and $c=1/2$, respectively.

Eq.\ (\ref{eq:Dc}) was derived in the scaling limit for the \emph{accessible} or \emph{external} hull of the clusters \cite{Grossman, Duplantier_Path}. For a finite system the external hull is
approximately produced by closing all fjords on the complete hull with a neck size of order unity.   The closing of the fjords reduces the dimension for percolation and the Ising model from $7/4$ and $11/8$ for the complete hull to $4/3$ and $5/3$ for the external hull, respectively \cite{Duplantier_Potts}.  The theory derived for percolation compares well with previous simulation results \cite{Meakin}.  However, these simulations were limited to relatively small systems, $\approx$ $10^5$ sites and 
did not probe deep into the fractal surface.

The authors of Ref.\ \cite{Meakin} used the method mentioned in the first paragraph:  a large number of walkers were released far away from the cluster and allowed to diffuse until they were absorbed on the hull.  This method is only able to measure the growth probabilities to an accuracy of $\approx 10^{-10}$, while percolation clusters with $10^5$ sites can have regions of the hull with probabilities smaller than $10^{-100}$.  Although these regions do not contribute to $D(q)$ for large $q$, they have a significant impact on $D(q)$ for small and negative $q$.  The algorithm of this paper can measure probabilities down to $10^{-300}$
which completely samples lattice systems with $\sim 10^4$ hull sites.  We have applied
this method on systems as large as $\sim 10^6$ hull sites.  In this paper we consider only percolation and Ising clusters though our method is quite general and can also be applied to off-lattice clusters.

\begin{figure}
\includegraphics[width=.40\textwidth]{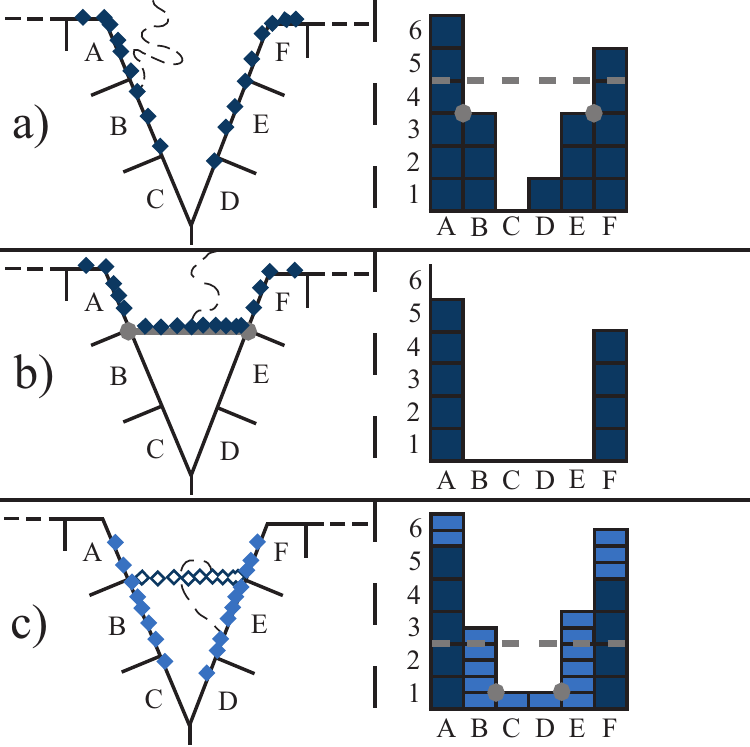}
\caption{\label{fig:FigSP} The signpost algorithm.  Left: the hull of the cluster (line) and the adsorbed random walkers (dots). Right: the number of walkers that hit each of the sections of the hull (A-F). Probe step, a) $18$ random walkers are released and adsorb onto the hull.  The histogram shows the distribution of hits with a dashed line representing the probability threshold, $4$. The large dots on the histogram show where the two ends of the signpost should be placed.  b) $18$ more walkers are released and adsorb onto the hull and the signpost (grey), exactly $9$ walkers on each.  The histogram shows the probability on the hull.  c) The next probe step: $18$ walkers   are released from the locations along the signpost where walkers in the step b) landed (open circles).  These walkers have half the weight as the ones released in parts a) and b).  The dashed line and 
 circles on the histogram show the new threshold and signpost end points. Color online}
\end{figure}

In order to treat very rare events, we use a \emph{biased sampling} of random walkers, in which we keep track of the ``lucky" random walkers that penetrate deep into the fjords of the hull.  The method has three steps: In the first iteration, $N$ random walkers with equal weight are released from outside the cluster and allowed to diffuse until they touch the hull. Each walker has weight $1/N$. The weights of the walkers are temporarily added to the probability distribution of the site where they land.  This step probes the cluster to find where the regions of small measure lie.  The hull sites that bound regions below a threshold probability (say $0.1$ to start out) are used as the end points of an absorbing line, or \emph{signposts}.   Then, the probability added in the first step is removed from the distribution and  $N$ more random walkers are released in a similar fashion as the first step.  These walkers can either diffuse onto the hull or a signpost line.  The weights of the walkers that touch the hull in this step are permanently added to the probability distribution. After all walkers have been absorbed, the signpost lines are removed. 

\begin{figure}
\includegraphics[width=.40\textwidth]{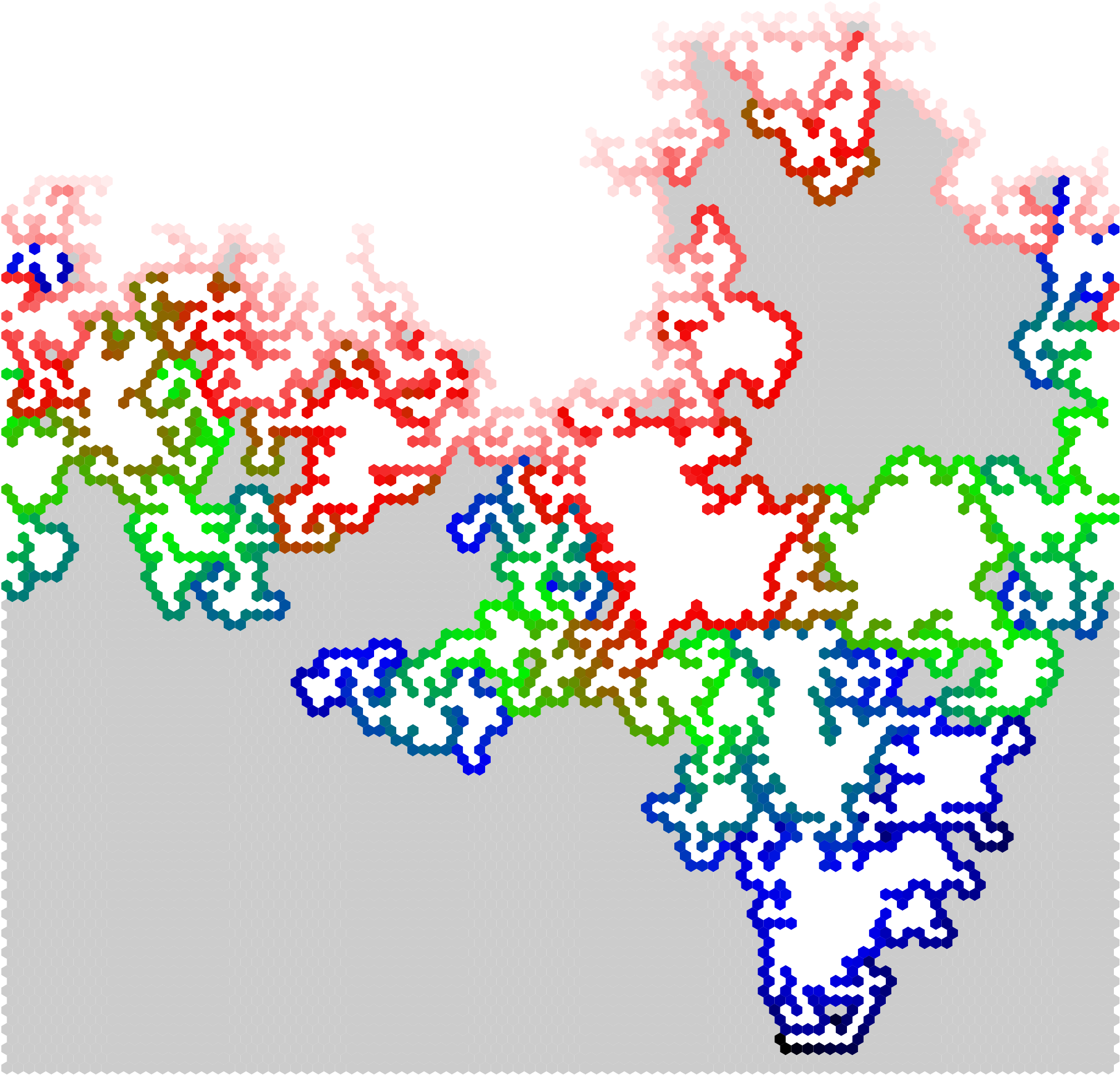}
\caption{\label{fig:Picture} The results of our simulation on the complete hull of a percolation cluster, where the harmonic measure goes from high to low though the colors red, green, then blue.  The scale is $\approx 50$ orders of magnitude. Sites outside cluster are white, and inside, grey. Color online.
}
\end{figure}

In the next iteration, the probe step is repeated with $N$ walkers \emph{released from the locations along the signpost lines} where walkers absorbed in the previous iteration. The threshold for small probability is reduced by a constant factor, e.g., 10. These walkers will have their weight reduced by the percentage of the walkers in the last step that touched a signpost line. 
This method is similar to methods used in chemical physics to numerically obtain reaction coordinate information \cite{Chem_1}. Fig.~\ref{fig:Picture} shows the harmonic measure of the external hull of a percolation cluster obtained using this method.

Our simulations are performed on a periodic triangular lattice with height $h$ and width $w$
such that $h=100w$, so that we obtain clusters that wrap around in width but not in height.  One ambiguity which must be resolved is the definition of random walkers touching the hull.  Here, we interpret this as the walker hopping \emph{onto} a hull site.  

The percolation clusters are grown using the Leath algorithm \cite{Leath}, with $p$ equal to the site threshold  for the triangular lattice, $p_c=1/2$. 
If a given cluster spans the width of the system, the top hull of the cluster is found using a simple border walking algorithm
related to the method of generating percolation cluster perimeters by random walks \cite{Ziff84}.
The list of complete hull sites of the cluster is then used in the signpost method to obtain the harmonic measure.  If the topmost vacant sites bordering the cluster are used instead of the occupied sites as the adsorbing sites, one obtains the external hull.

\begin{figure}
\includegraphics[width=.45\textwidth]{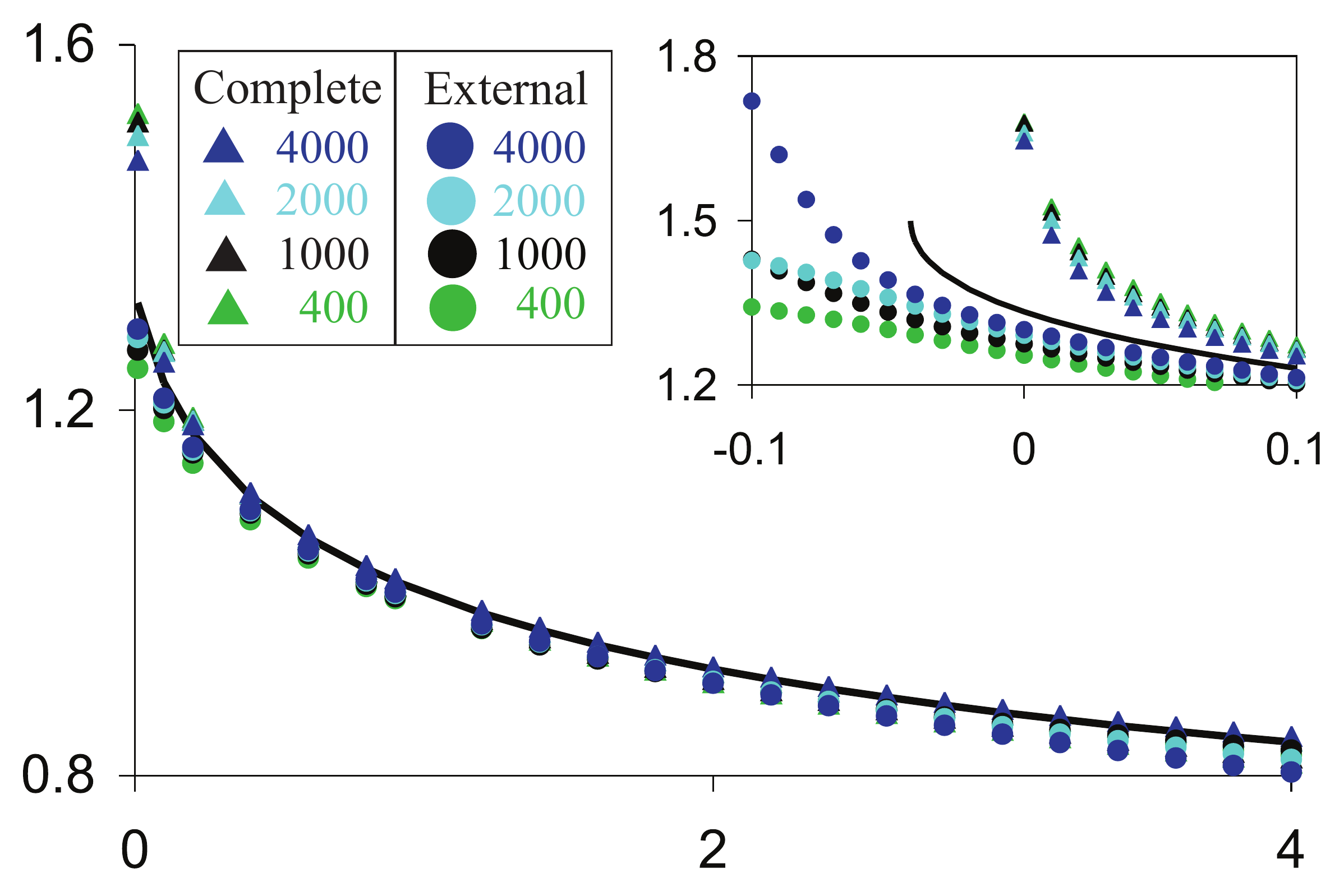}
\caption{\label{fig:Fig1} $D(q)$ vs. $q$ for the complete and external hulls of percolation clusters for four different widths compared to theory, Eq.\ (\ref{eq:Dc}) \cite{Duplantier_Per} (black line). Inset: small $q$ behavior of $D(q)$. Color online.}
\end{figure}

For the FK clusters on the Ising model, bonds are  placed between adjacent same-spin sites with probability $p_c = 1 - e^{-\beta J_c}$, where on the triangular lattice $p_c = 1 - 1/\sqrt{3}$.
  We use the Swendsen-Wang method \cite{Swendsen87} to equilibrate the system and simultaneously generate the FK clusters.  After the  system is sufficiently equilibrated, we attempt to find a spanning cluster.  These spanning bond clusters must be converted to site clusters if they are to be used with our algorithm. 
We do this by making another triangular lattice with half the lattice spacing, such that all bonds are centered on even sites in the new lattice.  Bonds are copied to the new lattice at the even sites on which they are centered.  Odd sites are added to the cluster if two adjacent bonds meet at the odd site.  Next, the perimeter-walk algorithm is used to record the locations of the hull sites;  then we use the signpost method to find the harmonic measure.  As in percolation, an external hull can be obtained, though for Ising clusters it requires the addition of artificial vacancies to all sites bordering the cluster.

The signpost method iteratively obtains smaller and smaller probabilities by reducing the weight of the random walkers released in each round, in our case by a factor of 10.  
We took the number of walkers, $N$ to depend on the system width, $w$. For example, for $w=400$ we use $N = 2\cdot 10^6$  and for $w=4000,$ $N= 2\cdot10^7$.  The signpost method is performed until all probabilities have been measured or until the minimum measurable probability $10^{-300}$ has been reached.  This minimum is close to the smallest value that can be stored in a double precision floating point number.  Significantly smaller values could be obtained by storing the log of the probability instead of the probability itself, though we have not implemented it.

The locations of the sites and their associated probabilities are then used to obtain $D(q)$ and the histogram of the probability distribution.  $D(q)$ is obtained by applying a linear fit to $\log Z_q $ in Eq.\ (\ref{eq:Zq}) versus $\log l$, where $l$ is the box length.  The fit was performed for a range of $l$ over which the function was linear. 
The histogram of the probability distribution is measured by using exponentially smaller bin sizes, e.g., the first box has size $1/2$, the next $1/4$, then $1/8$, etc. The exponent of the power-law is fit at different probabilities using $5$ points which roughly span an order-of-magnitude in probability.

\begin{figure}
\includegraphics[width=.45\textwidth]{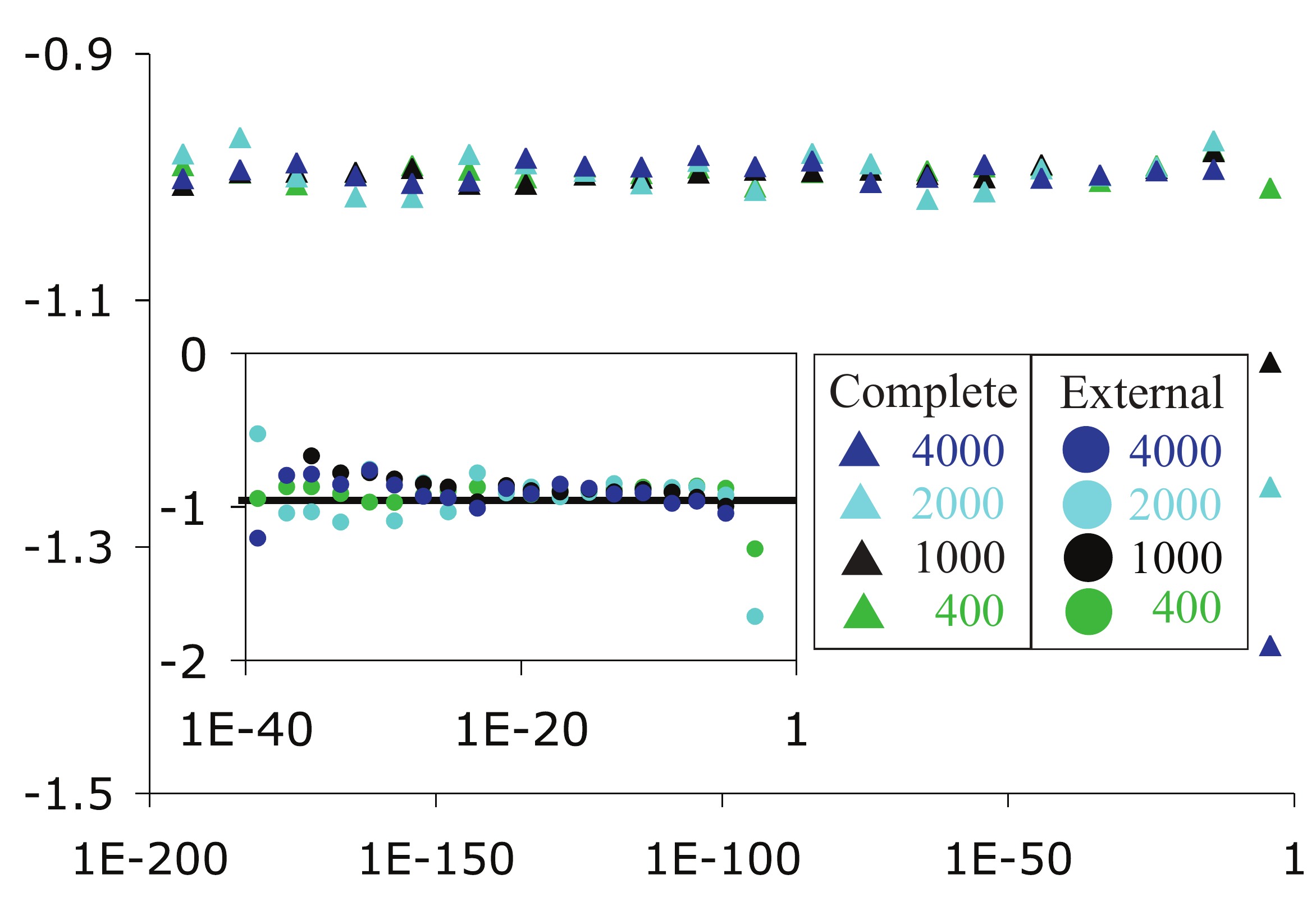}
\caption{\label{fig:Fig2} Exponent of the power-law fit to the probability distribution vs. probability for the complete hull of a percolation cluster for several different widths.  The inset shows a similar plot for the external hull with the associated theoretical prediction (black line) \cite{Duplantier_Per}. Color online}
\end{figure}

Simulations of percolation and Ising clusters were performed for a number of system widths.
Our results are for $w=400$, $1000$, $2000$, and $4000$, for which data was recorded for all systems. Small systems, $w=400$, have $\approx 5\cdot10^5$ hull sites in the cluster and large systems, $w=4000$, have $\approx 5\cdot10^6$ hull sites. $D(q)$ and the slope of the power-law fit to the probability distribution were obtained for the complete and external hulls of both percolation and the Ising clusters.

Fig.~\ref{fig:Fig1} shows a comparison between the results of the complete and external hulls of percolation clusters with the theory for the external hulls, Eq.\ (\ref{eq:Dc}).  There is good agreement among all three for large $q$, which is not surprising as the complete hull fjords contribute negligibly to $D(q)$.  For small $q$ there is significant disagreement between the complete hull and the theory as the two must approach different values for $D(0)$.  Previous simulations \cite{Grossman}   have shown that $D(0)$  increases with increasing width; however we see a peak at  a width of 100; see Fig. \ref{fig:Fig1}.  This is because there is a non-negligible fraction of the hull sites with probabilities less than $10^{-300}$ for large widths.  We expect for very large systems, if we are able to record all probabilities, that the complete hull $D(q)$ will be nearly identical to the theory for $q>0$, at $q=0$ there will be a jump to $D(0)=7/4$, and for $q<0$, $D(q)$ will be ill-defined (unbounded). In 
 comparison, for the external hull we see excellent agreement between the data and the theory (\ref{eq:Dc}) over the entire range of $D(q)$, especially for largest system sizes.

\begin{figure}
\includegraphics[width=.45\textwidth]{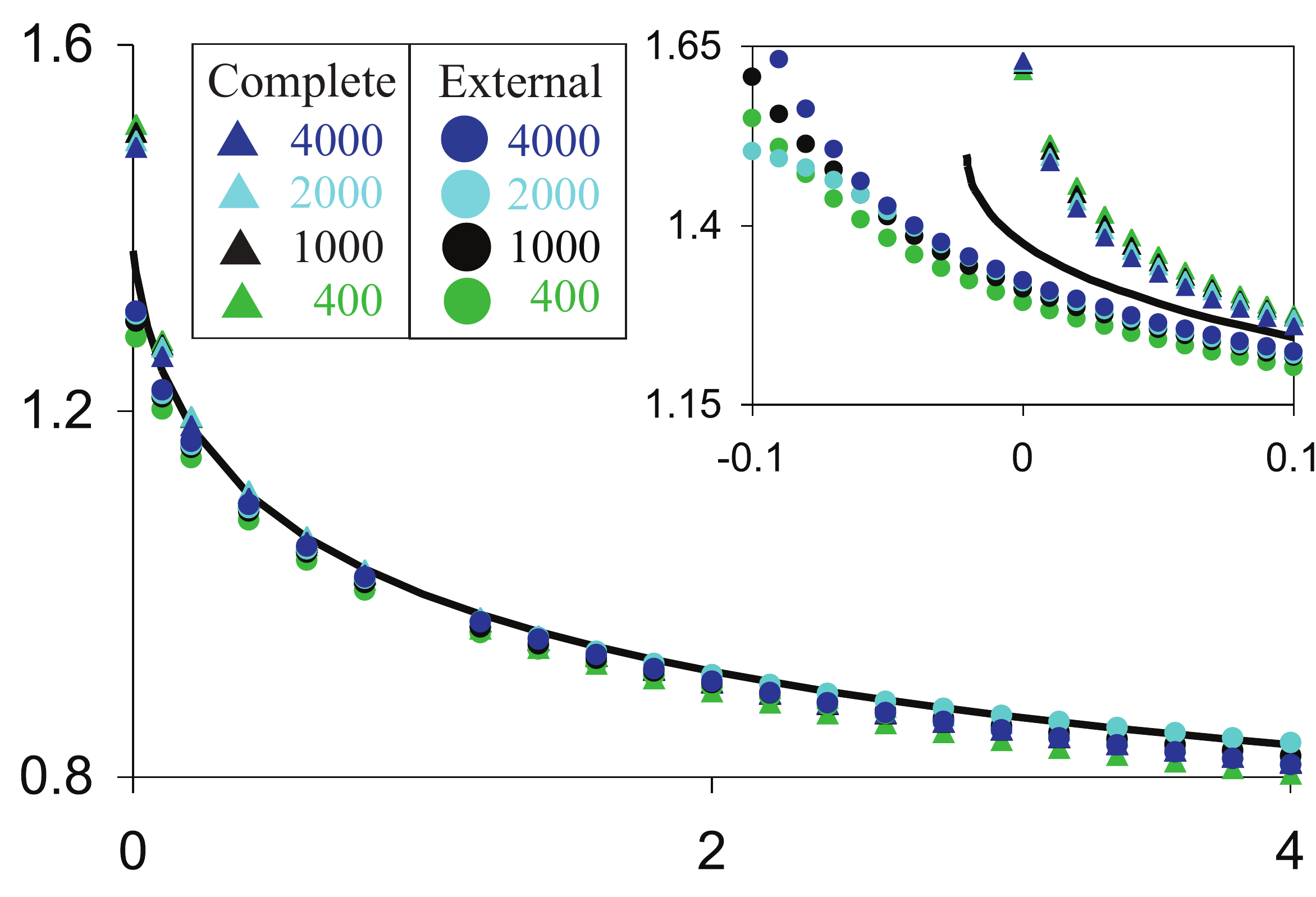}
\caption{\label{fig:Fig4} $D(q)$ vs. $q$ for the complete and external hulls of Ising clusters for four different widths compared to theory from Eq.\ (\ref{eq:Dc})  (black line). Inset: small $q$ behavior of $D(q)$. Color online.}
\end{figure}

The exponent of the power-law fit to the probability distribution for the complete and external hull (inset) are shown in Fig.~\ref{fig:Fig2}.  The exponent for the complete hull is $-0.996\pm 0.01$ over, incredibly, $50$ orders of magnitude!  We presume that the exact value of the exponent is $-1$, which implies that $D(q)$ is undefined for $q <0$.  Previous simulations \cite{Meakin} were unable to obtain this result because the smallest probability that could be measured, $\approx 10^{-10}$, is still in the transient regime.  The initial overshoot of the power for small systems corresponds to the probability distribution for the \emph{external} hull being picked up by the complete hull.  The power-law exponent is also obtained for the external hull, $-0.93\pm 0.05$, which is consistent  with the theoretical prediction of $-23/24 \simeq -0.958$.

Similar results were obtained for the Ising model.  Fig.~\ref{fig:Fig4} shows the comparison between the complete and external hulls of Ising clusters with the theory \cite{Duplantier_Potts} for $D(q)$.  As with percolation, there is good agreement with theory for large $q$ for both the complete and external hulls but significant disagreement at small $q$ for the complete hull, where the Eq.\ (\ref{eq:Dc}) does not apply.  The probability power-law exponents for the complete and external hull are, -0.997$\pm$0.012 and -0.920$\pm$0.048, respectively for the Ising model.  The complete hull exponent again points to $q=0$ as the discontinuity point for $D(q)$; the external hull exponent agrees roughly with theory.


In summary, we have described a method to obtain harmonic measures representing events of extremely low probability, which allowed us to probe the internal structure of percolation and Ising model cluster hulls, where we observe the asymptotic behavior $D(q) \sim q^{-1}$.    We are not aware of any theoretical prediction or explanation of this result.

In future work \cite{Adams09}, we plan to apply the continuous version of this algorithm  to obtain the harmonic measure for Diffusion Limited Aggregation (DLA) \cite{Witten81}  for which there are no exact results, though there are several  conjectures for the form of $D(q)$ for small and negative $q$ \cite{Jensen02}.  For the theory of DLA, the harmonic measure plays a central role, because it represents the growth probability at every point on the cluster at a given time.
The best current results for $D(	q)$ use iterative conformal maps \cite{Hastings98, Davidovitch99}, and are restricted to rather small clusters of $\approx 10^4$ sites. Our method will allow us to go to much larger sizes, $\approx 10^7$ sites. This is important because the slow crossover of some length scales in DLA \cite{Somfai99} suggests large clusters are necessary to approximate the scaling limit.  Our method could shed light on the internal structure of DLA for which little is known.

We thank A. Voter and B. Duplantier for very useful conversations. This work was supported in part by the National Science Foundation through DMS-0553487.

\bibliography{Draft_Bib}

\end{document}